# Primordial Black Holes

Jane H. MacGibbon
*Dept of Physics, University of North Florida, Jacksonville, FL, 32224 USA*

Tilan N. Ukwatta
*Space and Remote Sensing (ISR-2) and Physics Division (P-23), Los Alamos National Laboratory, Los Alamos, NM, 87544, USA*

J.T. Linnemann, S.S. Marinelli, D. Stump, K. Tollefson
*Dept of Physics and Astronomy, Michigan State University, East Lansing, MI, 48824 USA*

Primordial Black Holes (PBHs) are of interest in many cosmological contexts. PBHs lighter than about $10^{12}$ kg are predicted to be directly detectable by their Hawking radiation. This radiation should produce both a diffuse extragalactic gamma-ray background from the cosmologically-averaged distribution of PBHs and gamma-ray burst signals from individual light black holes. The Fermi, Milagro, Veritas, HESS and HAWC observatories, in combination with new burst recognition methodologies, offer the greatest sensitivity for the detection of such black holes or placing limits on their existence.

## 1. INTRODUCTION

A black hole (BH) is an object of classical gravity [1] whose mass $M_{BH}$ is contained within its Schwarzschild volume which has radius

$$r_{BH} = \frac{2GM_{BH}}{c^2} \quad (1)$$

[2]. Here $G$ is the universal gravitational constant, $c$ is the speed of light and we have assumed that the BH has negligible rotation and/or electric charge. (Extension in General Relativity to include rotation and/or electric charge is straightforward.) Because Eq (1) implies that the average density inside a black hole goes as $\rho_{BH} \propto M_{BH}/r_s^3 \propto M_{BH}^{-2}$, large mass black holes may be more easily produced than small mass black holes, at least in the present universe. In fact a $10^8 M_\odot$ black hole has the density of water. Today there is strong evidence for the existence of stellar mass black holes (formed as supernova remnants) and $10^6 M_\odot$-$10^{10} M_\odot$ supermassive black holes in most galactic centers. There is also mounting evidence for black holes with masses intermediate between stellar mass black holes and supermassive black holes.

'Primordial Black Hole' (PBH) refers to a black hole of any size formed in the early universe (where by 'early universe' we mean before the formation of the first stars). Possible PBH formation mechanisms include the collapse of overdense regions arising from primordial density inhomogeneities (such as occur in many Inflation models, in particular those with a blue, peaked or 'running index' spectrum), an epoch of low pressure (soft equation of state), or cosmological phase transitions; and mechanisms involving topological defects, such as cosmic strings oscillating into their Schwarzschild volume or the collapse of domain walls. (For a recent review of PBH formation mechanisms and limits see [3] and references therein.) In almost all scenarios, the PBH mass at the time of formation is roughly equal to, or smaller than, the cosmic horizon (or Hubble) mass $M_H \approx 10^{15}(t/10^{-23}s)$g. Thus the range of possible PBH initial masses is enormous – from the Planck mass for PBHs forming around the Planck time, to $10^5 M_\odot$ for PBHs forming around 1 s, or larger if forming later. Within a particular formation scenario, usually the PBHs are produced over a narrow initial mass range. An exception is scale-invariant cosmological primordial density perturbations which could produce PBHs over an extensive initial mass range with an initial mass spectrum of the form $dn/dM_i \propto M_i^{-\alpha}$ where $\alpha = 5/2$ for formation in the radiation era. Although scale-invariant density perturbations are not as well motivated in present cosmological models as they were a couple of decades ago, gamma-ray limits on the present cosmologically-averaged number density of PBHs were earlier derived assuming an $M_i^{-5/2}$ initial mass function.

The formation constraints on PBHs inform us about cosmology. The PBHs themselves may also produce effects on cosmological scales. PBHs surviving today should behave as cold dark matter (CDM). (In fact, present limits allow $10^{17} - 10^{26}$g PBHs to contribute all of $\Omega_{CDM}$ [3].) Like other CDM, PBHs should cluster in galactic haloes. They may also enhance the clustering of other dark matter, for example in WIMP and Ultra Compact Massive Halo scenarios. If a stable state such as a Planck mass relic remains after low mass PBHs have expired, the relics themselves are CDM candidates. PBHs may have played a role in the development of cosmological entropy, baryogenesis, the reionization of Universe in earlier epochs and producing observable annihilation lines. Very large PBHs may influence large scale structure development, seed SMBHs, or generate observable cosmic x-rays in their accretion disks.

The number of PBHs formed with initial masses of $10^9 - 10^{43}$g have been constrained primarily by primordial nucleosynthesis, cosmic microwave background (CMB) anisotropies, MACHO searches and, in the case of $M_{BH} \lesssim 10^{17}$g BHs, the search for Hawking radiation. Hawking radiation constraints derived from the 100 MeV extragalactic gamma-ray background and Galactic gamma-ray, $e^+$, $e^-$ and anti-proton backgrounds place an upper limit on the background distribution of $M_{BH} \approx 5 \times 10^{14}$g PBHs of roughly $\Omega_{PBH} \lesssim 10^{-9}$. Direct searches for the final gamma-ray burst of Hawking radiation from an expiring PBH allow us to directly constrain the local number density of $M_{BH} \approx 5 \times 10^{14}$g PBHs and much lighter BHs.

## 2. BLACK HOLE BURSTS

### 2.1. Black Hole Thermodynamics

The work by Hawking and Beckenstein in the 1970's on extending the Laws of Classical Thermodynamics to include black holes (i.e. Classical Gravitation) resulted in the recognition of the Hawking (Gravitational) temperature $T_{BH}$

$$kT_{BH} = \frac{\hbar c^3}{8\pi G M_{BH}} = 1.06 \left(\frac{M_{BH}}{10^{13}\text{g}}\right)^{-1} \text{GeV} \quad (2)$$

where $k$ and $\hbar$ are the Boltzmann and reduced Planck constants, respectively [4]. An $M_\odot$ black hole has a temperature of $10^{-7}$ K; a $10^{25}$ g black hole has the same temperature as the present CMB; and a $10^{11}$ g black hole has a temperature of ~ 100 GeV. Hawking also derived the thermal flux radiating from a black hole of temperature $T_{BH}$ to be

$$\frac{d^2 N_s}{dt dQ} = \frac{\Gamma_s}{2\pi\hbar}\left[\exp\left[\frac{Q}{kT_{BH}}\right] - (-1)^{2s}\right]^{-1} \quad (3)$$

per particle degree of freedom where $Q$ is the energy of the Hawking-radiated particle, $s$ is the particle spin and $\Gamma_s$ is the absorption probability [5]. In the geometric optics (short-wavelength) limit, $\Gamma_s \approx 27 G^2 M_{BH}^2 Q^2 / \hbar^2 c^6$. Strictly Eqs (2) and (3) apply for a non-rotating, non-electrically charged black hole. Extension to a black hole with angular momentum and/or electric field is straightforward but because a small black hole emits its angular momentum and electric charge quickly [5] compared to cosmological timescales we will assume PBHs surviving today have negligible angular momentum and electric field.

In the standard (MacGibbon-Webber) emission picture, a black hole should directly Hawking-radiate those particles which appear non-composite compared to the wavelength of the radiated energy (or equivalently the black hole size) at a given $T_{BH}$ [6]. In order of increasing $T_{BH}$, as $T_{BH}$ surpasses successive particle rest mass thresholds, the black hole initially directly emits photons (and gravitons), then neutrinos, electrons, muons and eventually direct pions. Once $T_{BH} \gtrsim \Lambda_{QCD} \approx 200 - 300$ MeV, the QCD confinement scale, the black hole should directly Hawking-radiate, not pions which are now composite at such temperatures, but quarks and gluons. Analogous to QCD jet behaviour in accelerators, the quarks and gluons will subsequently shower and hadronize into the astrophysically stable species $\gamma, \nu, p, \bar{p}, e^-$ and $e^+$ as they stream away from the black hole. Because of the large number of degrees of freedom for the fundamental QCD particles, the instantaneous emission spectra from $T_{BH} > \Lambda_{QCD}$ black holes are dominated by the component produced by the decay of the Hawking-radiated QCD particles. The instantaneous photon flux from a $T_{BH} > \Lambda_{QCD}$ black hole is dominated by this secondary QCD photon component while the directly Hawking-radiated photons contribute, at a given $T_{BH}$, significantly only at the highest energies. For $T_{BH} = 0.3 - 100$ GeV black holes, the total instantaneous fluxes of the final-state stable particles are

$$\dot{N}_{p\bar{p}} \approx 2.1(\pm 0.4) \times 10^{23} \left[\frac{T_{BH}}{\text{GeV}}\right]^{1.6 \pm 0.1} \text{s}^{-1}$$

$$\dot{N}_{e^\pm} \approx 2.0(\pm 0.6) \times 10^{24} \left[\frac{T_{BH}}{\text{GeV}}\right]^{1.6 \pm 0.1} \text{s}^{-1}$$

$$\dot{N}_{\gamma} \approx 2.2(\pm 0.7) \times 10^{24} \left[\frac{T_{BH}}{\text{GeV}}\right]^{1.6 \pm 0.1} \text{s}^{-1}$$

$$\dot{N}_{\nu\bar{\nu}} \approx 5.6(\pm 1.7) \times 10^{24} \left[\frac{T_{BH}}{\text{GeV}}\right]^{1.6 \pm 0.1} \text{s}^{-1}$$

And the average energies of the fluxes scale as roughly $T_{BH}^{0.5}$, not as $T_{BH}$ (as for the directly Hawking-radiated components) [6]. Thus, even very high temperature black holes will produce significant fluxes of final state particles which have energies around 100 MeV – 1 TeV.

As the black hole Hawking-radiates, its mass is carried off by the mass-energy of the emitted particles. The black holes mass loss rate is thus

$$\dot{M}_{BH} \approx -5.34 \times 10^{25} f(M_{BH})(M_{BH}/\text{g})^{-2} \text{g s}^{-1} \quad (4)$$

where the weight $f(M_{BH})$ accounts for the total number of directly emitted states and is normalized to unity for $M_{BH} \gg 10^{17}$g black holes which emit only photons and the three neutrino species. The relativistic contributions to $f(M_{BH})$ per particle degree of freedom are $f_{s=0} = 0.267$, $f_{s=1/2} = 0.147$ (uncharged), $f_{s=1/2} = 0.142$ (charge $e^\pm$), $f_{s=1} = 0.060$, $f_{s=3/2} = 0.020$, and $f_{s=2} = 0.007$ [7]. For a $T_{BH} \approx 50$ GeV black hole emitting all experimentally-confirmed Standard Model degrees of freedom including the 125 GeV Higgs boson, $f(M_{BH}) \approx 15$.

Integrating Eq (4), the remaining evaporation lifetime of an $M_i$ black hole is then

$$\tau_{evap} \approx 6.24 \times 10^{-27} f(M_i)^{-1} (M_i/\text{g})^3 \text{s}. \quad (5)$$

The mass of a PBH whose evaporation lifetime equals the age of the universe is $M_* \approx 5.00(\pm 0.04) \times 10^{14}$g [8].

Comparison of the observed diffuse extragalactic gamma-ray background around 100 MeV with the gamma-ray background that would be produced by a cosmological distribution of $M_* \approx 5 \times 10^{14}$g PBHs places the strictest limit on an cosmologically-averaged distribution of $M_*$ PBHs. The limit, updated in 2010 using the Fermi LAT data, is $\Omega_{PBH}(M_*) \lesssim 5 \times 10^{-10}$ [3]. (This $M_*$ limit is stricter and more robust than the limits on the cosmological distribution of PBHs of any other mass derived by this or any other method.) Because PBHs should behave as CDM, however, they should not be uniformly distributed throughout the universe but should cluster in galactic halos (and possibly also on smaller scales). Assuming PBH clustering in the Galactic halo, the local number density of PBHs should be enhanced by a factor of $\eta_{local} \sim 2 \times 10^5 (\Omega_{halo}/0.1)^{-1}$ where $\Omega_{halo}$ is the cosmological density parameter associated with galactic halos [9]. Clustering in the Galaxy leads to the possibility that PBHs are contributing to the Galactic halo gamma-ray background (as investigated by Wright using EGRET observations [10]), matter-antimatter interactions and microlensing events. Comparisons of the spectra from a Galactic distribution of PBHs with the observed Galactic antiproton and positron backgrounds around 100 MeV lead to limits on a Galactic distribution of $M_* \approx 5 \times 10^{14}$g PBHs which are similar or somewhat weaker than the extragalactic gamma-ray limit. These antiproton- and positron-derived limits, however, depend on the modeling of the propagation and leakage times of charged particles in the Galaxy and on the Galactic distribution of PBHs, and so are not as robust as the extragalactic 100 MeV gamma-ray limit on the cosmologically-averaged distribution of PBHs.

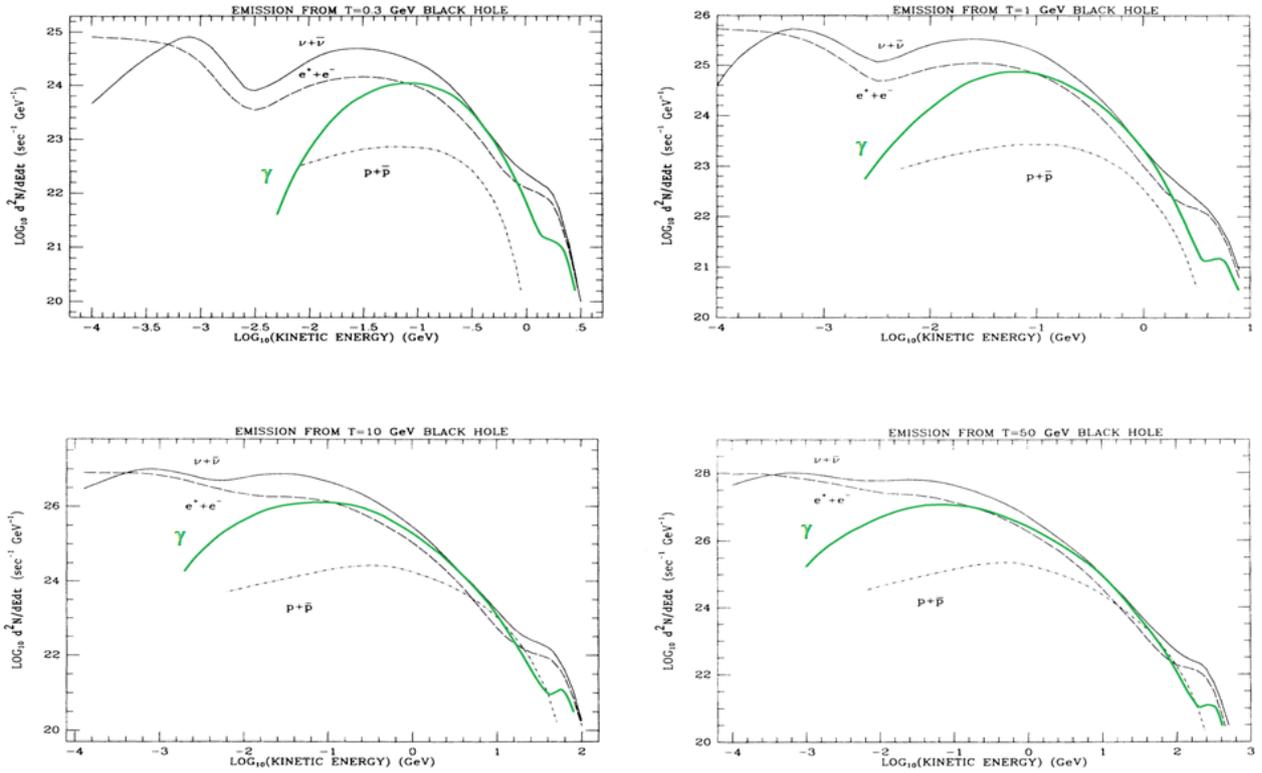

Figure 1: The instantaneous gamma-ray flux $d^2N/dtdE$ detectable by Fermi-LAT from $T_{BH} = 0.1 - 50$ GeV black holes [6]. For $T_{BH}$ in this range, the flux should remain approximately constant over the lifetime of the Fermi Observatory.

We note that the extragalactic and Galactic limits are derived using the black hole emission spectra integrated over both a distribution of PBHs and Galactic or cosmological timescales.

## 2.2 Signatures of Black Hole Bursts

Independently we can derive limits by directly searching for the present emission from an individual black hole. Equally importantly, we can predict the light curve that would be produced in a detector by an individual black hole and devise methodologies to distinguish the BH burst signal from other known gamma-ray source types. Burst searches are the direct method for detecting black hole Hawking radiation and do not depend on assumptions concerning the formation mechanism of the black hole. In fact, burst searches are equally searches for any local small black holes created in the present universe, as well as primordially-produced PBHs. Although there are no currently-fashionable theories predicting the production of such small black holes in the present Galaxy, we should not bias ourselves observationally against their possible existence, given the widespread acceptance of the existence in the Galaxy and beyond of stellar mass and higher mass black holes. We should investigate the black hole burst signature template so that we can recognize BH/PBH bursts if they are seen in a detector.

Let us now predict the black hole burst signature. Rewriting Eq (5), a black hole with temperature $T_{BH}$ has a remaining evaporation lifetime of

$$\tau_{evap} \approx 5.0 \times 10^{11} \left(\frac{f(T_{BH})}{15}\right)^{-1} \left(\frac{T_{BH}}{\text{GeV}}\right)^{-3} \text{ s.} \quad (6)$$

A $T_{BH} \approx 1$ GeV black hole has a remaining lifetime of ~ 16,000 yr; a $T_{BH} \approx 10$ GeV black hole has a remaining lifetime of ~ 20 yr; a $T_{BH} \approx 25$ GeV black hole has a remaining lifetime of ~ 1 yr; a $T_{BH} \approx 300$ GeV black hole has a remaining lifetime of ~ 1 hr; a $T_{BH} \approx 2$ TeV black hole has a remaining lifetime of ~ 100 s; and a $T_{BH} \approx 20$ TeV black hole has a remaining lifetime of ~ 100 ms.

As can be seen from Eqs (2) and (4), the $M_{BH} \ll M_*$ black hole's mass quickly decreases as it radiates and its temperature increases at an accelerating pace. Recall that the photons produced from the decays of the directly Hawking-radiated QCD particles dominate the net instantaneous photon flux from a $T_{BH} > \Lambda_{QCD}$ black hole and have an average energy that scales as roughly $T_{BH}^{0.5}$, not as $T_{BH}$. Thus substantial numbers of 100 MeV – 10 TeV photons will be produced even during the final explosive stage of the black hole's evaporative lifetime.

With respect to detecting gamma-ray black hole bursts with the Fermi Observatory, there are 3 cases of BH signals that we need to consider:

Case (i) The gamma-ray spectrum from a $3 \text{ MeV} < T_{BH} < 12 \text{ GeV}$ black hole will appear to be almost constant as a function of time over the lifetime of the Fermi Observatory. (Recall that the remaining evaporation lifetime of a $T_{BH} = 10$ GeV black hole is ~ 20 yrs.)

Case (ii) The gamma-ray spectrum from a $12 \text{ GeV} < T_{BH} < 50 \text{ GeV}$ black hole will evolve significantly as a function of time over the lifetime of the Fermi Observatory but almost all its gamma-ray flux arriving over that time will lie within the LAT detector's energy range, 20 MeV - 300 GeV. (Recall that the remaining evaporation lifetime of a $T_{BH} = 50$ GeV black hole is ~ 50 days.)

Case (iii) The gamma-ray spectrum from a $T_{BH} > 50$ GeV black hole will be a quickly evolving burst with part of its flux arriving in the LAT energy range and significant flux at energies above the LAT range. In the final stages of burst evolution, the incoming flux will not be resolvable as a function of time and the time-integrated flux will be deposited in one time interval in the detector. (Recall that the remaining evaporation lifetime of a $T_{BH} = 170$ TeV black hole is ~ 100 μs.)

In Figure 1, we show the instantaneous gamma-ray flux $d^2N/dtdE$ which would be seen by the LAT from $T_{BH} = 0.1 - 50$ GeV black holes [6], relevant to Cases (i) and (ii). For black holes with these temperatures the flux is dominated by the photons resulting from the Hawking-radiated QCD particles. The gamma-ray flux from a $T_{BH} = 20$ MeV black hole, which is below the threshold to emit a QCD component and whose photons are all directly Hawking-radiated, is shown in Figure 2 [6].

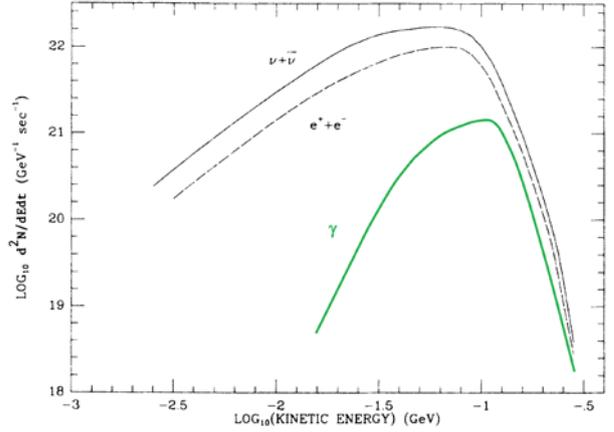

Figure 2: The instantaneous gamma-ray flux from a $T_{BH} = 20$ MeV black hole, which is below the threshold to emit a QCD component [6].

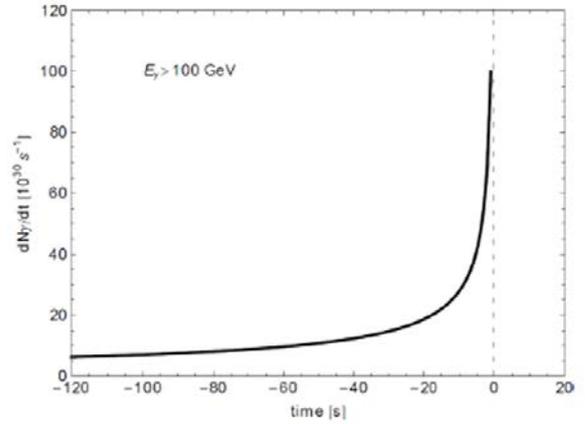

Figure 3: Preliminary calculation for the PBH burst light curve $dN/dt$ arriving in the detector with energy above a given threshold, here $E_\gamma = 100$ GeV.

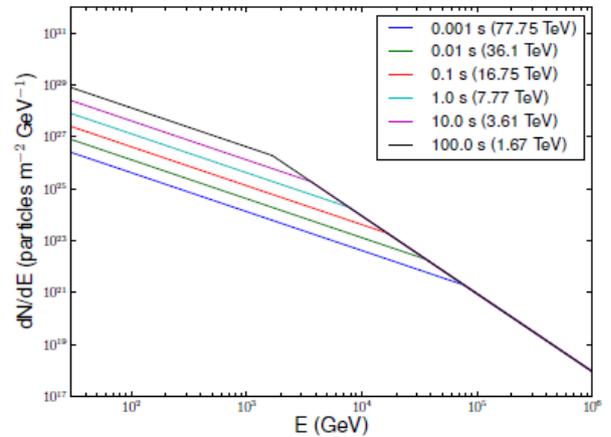

Figure 4: The gamma-ray spectrum $dN/dE$ time-integrated over various remaining black hole evaporation lifetimes [11].

For Case (iii), we show in Figure 3 our preliminary calculation for the PBH burst light curve, i.e. the number of photons arriving per unit time with energy above a given threshold. (In Figure 3, the energy threshold is taken to be $E_\gamma = 100$ GeV). In Figure 4, we plot the gamma-ray spectrum time-integrated over various BH remaining evaporation lifetimes [11].

In Table 1, we list a number of distinguishing characteristics to discern a black hole burst from other known GRB source types. In particular, the BH burst will show a soft-to-hard (that is, low average energy to high average energy) time evolution and will be non-repeating. If it is bursting in free space, it should not be accompanied by an afterglow, but generation of an afterglow may be possible if the black hole is bursting in an ambient high density plasma or ambient high magnetic field.

Table 1: Differences between black hole burst signals and GRBs of known source types.

| Gamma-Ray Bursts (known GRB types) | BH Bursts |
|---|---|
| Detected at cosmological distances | Local, unlikely to be detected from beyond Galaxy |
| Most GRBs show hard-to-soft evolution | Hard-to-soft evolution expected |
| Hadrons not expected from GRBs | Accompanied by hadronic bursts which may be detectable if local |
| Gravitational wave signal expected | No accompanying gravitational wave signal |
| Time duration ranges from fractions of second to hours | Time duration of burst most likely 1-100 seconds |
| Fast Rise Exponential Decay (FRED) light curve | Exponential Rise Fast Fall (ERFF) light curve |
| X-ray, optical, radio afterglows expected | No multi-wavelength photon afterglows unless in exotic ambient environment |
| TeV emission unknown | TeV spectra predicted |
| Multi-peak time profile | Single-peak time profile |
| May be repeating | No burst repetition |

If no black hole bursts are observed by a detector, the null detection implies an upper limit on the local number density of small black holes. An amalgamation of recent limits and limits which would be set by null detection with HAWC are shown in Figure 5. As a general statement, the strongest limits have been set by searching for bursts of about $1 - 100$ s duration because the detector signal weakens for bursts of shorter duration and the background dampens signal recognition at longer duration. The advantages [12] of the Fermi Observatory, are that it is not background-limited, it has good angular and time resolution, a wide field of view and a low energy threshold, and it is anticipated to have a very long operational lifetime. Preliminary limits derived from a search of Fermi LAT data to date for pairs of photons with an arrival interval shorter than the time expected for a Poisson-distributed photon background give an upper limit of $2 \times 10^3 \mathrm{pc}^{-3}\mathrm{yr}^{-1}$ on BH bursts of $10^5$ s duration (corresponding to $T_{BH} \gtrsim 200$ GeV and $M_{BH} \lesssim 6 \times 10^{10}$ g) [13].

The $\dot{M}_{BH} \propto M_{BH}^{-2}$ dependence of Eq (4) means that, for any population of black holes that have masses today around some $M_{BH} \ll M_*$ (i.e. that have remaining lifetimes much less than the age of the universe), the number of black holes per mass interval around $M_{BH}$ today is

$$\frac{dn}{dM_{BH}} \propto M_{BH}^2 \qquad (7)$$

independent of the BH formation time, formation mechanism or spatial distribution [9]. For black holes recently created with mass $M_i \ll M_*$, the distribution (7) applies around the evolved mass $M_{BH}$ even if the initial mass distribution had initially been almost a delta function at $M_i$ (because in reality there is always some smearing of such a delta function).

In the case of PBHs with initial masses of $M_i \sim M_*$ created in the early universe, the distribution (7) applies today up to $M_{BH} \sim M_*$ but the mass distribution with which the PBHs were initially created would still apply above $M_*$ today because $M_i > M_*$ PBHs have lost little mass over the history of the universe. Therefore, using Eq (7), we can extrapolate the burst search limits to derive a limit on the number of $M_i \sim M_*$ PBHs created in the early universe. All of the BH burst search limits to date when extrapolated up to $M_i \sim M_*$ correspond to limits on the cosmologically-averaged number density of $M_*$ PBHs which are weaker than the limit derived from the 100 MeV extragalactic gamma-ray background. For reasonable values of the enhancement due to CDM clustering in the Galaxy, the 100 MeV extragalactic limit on the cosmologically-averaged number density of $M_*$ PBHs corresponds to a local BH burst limit of $\sim 10 \mathrm{pc}^{-3}\mathrm{yr}^{-1}$.

It should be noted, however, that the BH burst search limits are robust limits on the number density of small black holes close to Earth, regardless of their formation epoch or formation mechanism. Such black holes, if they are observed, are not necessarily the evolved state of $M_i \sim M_*$ PBHs formed in the early universe. Also, the assumptions concerning the clustering or spatial distribution of local BHs/PBHs used in the analysis may be incorrect, making detection in a given scenario more or less likely.

## 2.3 Further Comments on the Black Hole Burst Spectra

In the above analysis, the black hole is assumed to Hawking-radiate only the experimentally-confirmed fundamental particle species of the Standard Model of particle physics. If further fundamental modes beyond the Standard Model exist, the extra modes may enhance both the instantaneous flux from the black hole and the rate at which $M_{BH}$ decreases and $T_{BH}$ increase. This will shorten the black hole's remaining evaporation lifetime and the duration of the final burst. If new fundamental modes appear only at temperatures well above 100 TeV, the overall effect on the predicted observable spectra is, most likely, negligible. A significant number of new fundamental modes at lower energies are postulated, though, in some extensions to the Standard Model but it is expected that the weighting factor in Eqs [4] – [6] remains of order $f(M_{BH}) \lesssim 100$ [7]. For example in Supersymmetry models, each $s = 1/2$ fundamental particle has an $s = 0$ superpartner and each $s = 1$ fundamental particle has an $s = 1/2$ superpartner, giving $f(M_{BH}) \lesssim 45$. The accompanying enhancement to the instantaneous flux and $dN/dE$ spectra at a particular energy would depend on the actual decay processes of the new modes.

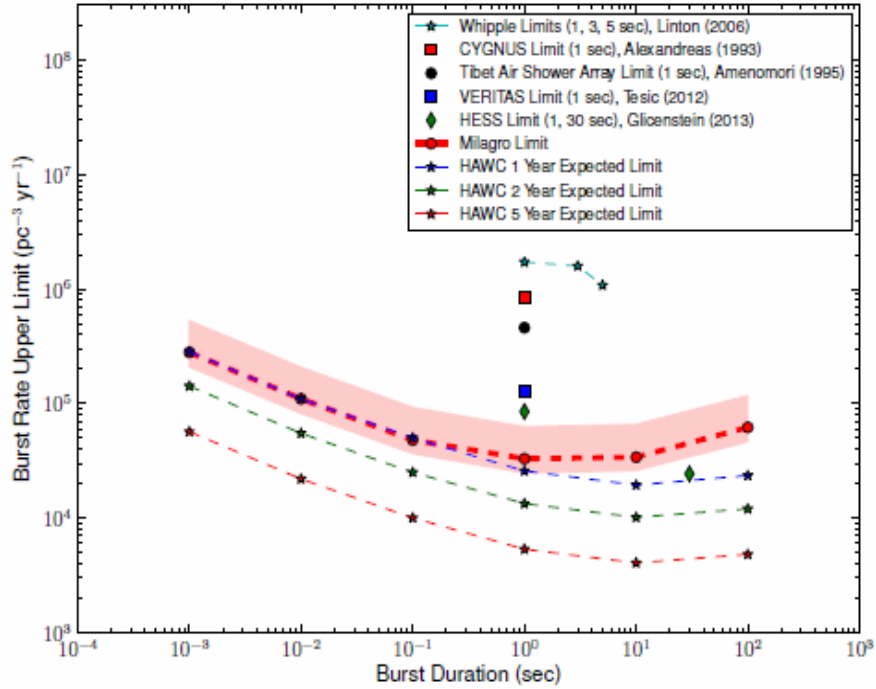

Figure 5: Limits on the local number density of black hole bursts in $pc^{-3}yr^{-1}$ set by null-detection in previous burst searches, together with projected limits which would be set by null-detection at the HAWC Observatory [11].

A number of PBH burst scenarios invoking significant self-interaction of the Hawking-radiated particles in the vicinity of the black hole after emission have been proposed. If such self-interaction did occur after emission, it would not change the remaining evaporation lifetime but would decrease the average energy of the photons arriving at the detector, i.e. decrease the expected observable spectra at high energies and increase the spectra at low energies [14]. Such photosphere models have recently been re-analyzed in detail and it has been strongly argued that the conditions for photosphere or quark-gluon plasma development are not met in the vicinity of the evaporating black hole [8]. Specifically, the time interval between successive Hawking emissions, the damping of Hawking emission and the limited amount of energy per emission near a species' rest mass threshold, and the Lorentz-transformed distance over which a scattered particle becomes 'on-shell' are such as to prevent the Hawking-radiated particles undergoing a significant number of QED or QCD interactions in the neighbourhood of the black hole. Hagedorn models [15] which invoke an exponential increase in fundamental hadronic states around a limiting temperature of $T_{BH} \sim m_\pi$, and which would produce a more detectable BH burst signal peaking at lower photon energy, are inconsistent both with accelerator experiments at these and higher energies (which confirm the quark model interpretation) and with the gravitational definition of $T_{BH}$ whose evolution is determined by the BH mass-energy loss rate. Hagedorn-type behaviour which may occur at extremely high energies in string theories would have negligible effect on the BH burst signal.

Although photospheres produced by intrinsic self-interaction of the radiated particles in the vicinity of a stand-alone BH appear to be ruled out, it may be possible to produce a non-intrinsic photosphere or distortion of the burst signal if the BH is embedded in an ambient high density plasma or strong magnetic field. Such scenarios have not yet been modeled in detail. The standard emission model BH gamma-ray spectra also do not yet incorporate the recently-recognized inner bremsstrahlung (single-vertex bremsstrahlung) component which is expected to dominate the directly Hawking-radiated photon component below about 50 MeV [16].

## 3. SUMMARY

There is strong motivation for investigating the possibility of detecting black hole burst signals. Detection of an evaporating black hole burst would be definitive experimental proof of the amalgamation of classical gravity with classical and quantum thermodynamics, pioneered by Hawking and Bekenstein. Equally importantly, the final stages of the evaporation process would open a direct observational window into particle physics at energies higher than can ever be achieved with terrestrial accelerators. For example, the black hole evaporation rate will be significantly increased if the supersymmetry modes exist. Details of the final stage of the BH burst may give insight into a quantum aspect of gravitation. Deviations of the BH burst signature from the predicted standard emission model spectra could also be used a probe of ambient extreme astrophysical environments. Detection or non-detection of PBHs give important constraints on the conditions in the early universe, in particular the amplitude and spectral index of initial density perturbations on smaller scales than are probed by the cosmic microwave background measurements. Thus, even if there is null-detection of BH bursts, there is strong motivation for improving the search limits and the implied upper limits on the number density of PBHs.

Updated detailed modeling of the BH burst signal that could be observed by the Fermi Observatory and exploration of new search methodologies is currently ongoing.

## Acknowledgments

JHM wishes to thank the organizers of the 5th Fermi Symposium for their hospitality.